# Visualizing Marden's theorem with Scilab

Klaus Rohe, 85625 Glonn, Germany, email: klaus-rohe@t-online.de

## Abstract

A theorem which is named after the American Mathematician Moris Marden states a very surprising and interesting fact concerning the relationship between the points of a triangle in the complex plane and the zeros of two complex polynomials related to this triangle.

> **Suppose the zeroes $z_1$, $z_2$, and $z_3$ of a third-degree polynomial p(z) are non-collinear. There is a unique ellipse inscribed in the triangle with vertices $z_1$, $z_2$, $z_3$ and tangent to the sides at their midpoints: the *Steiner in-ellipse*. The foci of that ellipse are the zeroes of the derivative p'(z).** (Wikipedia contributors, "Marden's theorem", http://en.wikipedia.org/wiki/Marden%27s_theorem).

This document describes how Scilab, a popular and powerful open source alternative to MATLAB, can be used to visualize the above stated theorem for arbitrary complex numbers $z_1$, $z_2$, and $z_3$ which are not collinear. It is further demonstrated how the equations of the Steiner ellipses of a triangle in the complex plane can be calculated and plotted by applying this theorem.

# Content

# 1 Introduction

The following theorem which is named after the American Mathematician Moris Marden states a very surprising and interesting fact concerning the relationship between the points of a triangle in the complex plane and the zeros of two complex polynomials related to that triangle. The theorem is quoted from reference [7]:

> **Suppose the zeroes $z_1$, $z_2$, and $z_3$ of a third-degree polynomial p(z) are non-collinear. There is a unique ellipse inscribed in the triangle with vertices $z_1$, $z_2$, $z_3$ and tangent to the sides at their midpoints: the *Steiner in-ellipse*. The foci of that ellipse are the zeroes of the derivative p'(z).**

An elementary proof of the theorem is given in [3]. It is also mentioned in this reference that the theorem has been published by J. Siebeck in 1864 in the Journal für die Reine und Angewandte Mathematik, Vol. 64, page 175- 182. Another source where the essential propositions of the above theorem are stated is reference [2], page 155.

In the following it is shown how, Scilab[1] a popular and powerful open source alternative to MATLAB, can be applied to visualize the above stated theorem for arbitrary complex numbers $z_1$, $z_2$, and $z_3$. The release of Scilab which will be used is the current version 5.5.1. This is not intended to be an introduction to Scilab rather it should be regarded as a case study which shows by example how Scilab can be applied to solve a given mathematical problem.

Before we are going to write the code to visualize this theorem with Scilab we have to carry out some preliminary mathematical tasks:

1) We have to compute the zeros of $p'(z)$ to get the actual values $z_{F_1}$ and $z_{F_2}$ for the foci of the ellipse[2].
2) From $z_1$, $z_2$, $z_3$, $z_{F_1}$ and $z_{F_2}$ we have to determine the equation of the ellipse which in general has the structure $A * x^2 + B * x * y + C * y^2 + D * x + E * y + F = 0$. This means we have to calculate the concrete values of the six coefficients $A$ to $F$.

To understand the following mathematical derivations it is the assumed that the reader is familiar with complex numbers and the basic features of conics and especially the ellipse.

---

[1] See URLs in reference [6], c) and d) are a very good tutorial like introductions to Scilab.
[2] In this paper I will not distinguish between a complex number z and the point which represents it in the complex plane also called the Gauss or Argand plane.

# 2 Preliminary mathematical topics

## 2.1 Computing the foci of the ellipse

The complex polynomial $p(z)$ with roots $z_1$, $z_2$, $z_3$ can be written as

$p(z) = (z - z_1) * (z - z_2) * (z - z_3)$ which can be expanded to

$p(z) = z^3 + (z_1 + z_2 + z_3) * z^2 + (z_1 * z_2 + z_1 * z_3 + z_2 * z_3) * z - z_1 z_2 z_3$

The first derivative of this polynomial is

$\frac{dp}{dz} = p'(z) = 3 * z^2 + 2 * (z_1 + z_2 + z_3) * z + z_1 * z_2 + z_1 * z_3 + z_2 * z_3$

To compute the zeros of the derivative we have to solve the quadratic equation

**(1)** $3 * z^2 + 2 * (z_1 + z_2 + z_3) * z + z_1 * z_2 + z_1 * z_3 + z_2 * z_3 = 0$

If we define $a = (z_1 + z_2 + z_3)$ and $b = z_1 * z_2 + z_1 * z_3 + z_2 * z_3$ equation (1) becomes

$z^2 + \frac{2*a}{3} * z + b = 0$. This equation has the solutions

$$(1) z_{F_1} = \frac{1}{3} * \left(a + \sqrt{a^2 - 3 * b}\right) = \frac{a}{3} + \sqrt{a^2 - \frac{b}{3}}$$

$$(2) z_{F_2} = \frac{1}{3} * \left(a - \sqrt{a^2 - 3 * b}\right) = \frac{a}{3} - \sqrt{a^2 - \frac{b}{3}}$$

Substituting $a$ and $b$[3] in the above equations and expanding the resulting terms gives the following results

$$z_{F_1} = \frac{1}{3} * \left(z_1 + z_2 + z_3 + \sqrt{(z_1^2 + z_2^2 + z_2^2) - (z_1 * z_2 + z_1 * z_3 + z_2 * z_3)}\right)$$

$$z_{F_2} = \frac{1}{3} * \left(z_1 + z_2 + z_3 - \sqrt{(z_1^2 + z_2^2 + z_2^2) - (z_1 * z_2 + z_1 * z_3 + z_2 * z_3)}\right)$$

The complex numbers $z_{F_1}$ and $z_{F_2}$ represent the foci of the ellipse we are looking for. To specify this ellipse uniquely we need at least one point on the ellipse (see appendix(I)). As Marden's theorem states "…***and tangent to the sides at their midpoints…***" it follows that the complex numbers $z_{E_1} = \frac{z_1 + z_2}{2}$, $z_{E_2} = \frac{z_1 + z_3}{2}$ and $z_{E_3} = \frac{z_2 + z_3}{2}$ are all part of this ellipse.

We now consider the special case that the quadratic equation (1) has one solution only which happens if $(z_1^2 + z_2^2 + z_3^2) - (z_1 * z_2 + z_1 * z_3 + z_2 * z_3) = 0$. It follows that the identity $z_1^2 + z_2^2 + z_3^2 = z_1 * z_2 + z_1 * z_3 + z_2 * z_3$ holds. According to reference [4], page 76 this is equivalent to the fact that the numbers $z_1$, $z_2$, $z_3$ are the vertices of an equilateral triangle in the complex plane. Because of $z_{F_1} = z_{F_2}$ the ellipse is a circle. It is the in-circle of this equilateral triangle.

---

[3] Not to be confused with the semi major and minor axes of the ellipse which are usually named with the same characters.

So the ellipse we are looking for is uniquely characterized by the following complex numbers:

**The two foci:**

$$z_{F_1} = \frac{1}{3} * \left( z_1 + z_2 + z_3 + \sqrt{(z_1^2 + z_2^2 + z_2^2) - (z_1 * z_2 + z_1 * z_3 + z_2 * z_3)} \right)$$

$$z_{F_2} = \frac{1}{3} * \left( z_1 + z_2 + z_3 - \sqrt{(z_1^2 + z_2^2 + z_2^2) - (z_1 * z_2 + z_1 * z_3 + z_2 * z_3)} \right)$$

**Three points on the ellipse:**

$$z_{E_1} = \frac{z_1 + z_2}{2}$$

$$z_{E_2} = \frac{z_1 + z_3}{2}$$

$$z_{E_3} = \frac{z_2 + z_3}{2}$$

**Box 1 Formulas to compute the foci and three points on the Steiner in-ellipse**

The center of the ellipse is $z_0 = \frac{z_{F_1} + z_{F_2}}{2} = \frac{\frac{2}{3}*(z_1 + z_2 + z_3)}{2} = \frac{z_1 + z_2 + z_3}{3}$. So $z_0$ is also the centroid of the triangle with vertices $z_1$, $z_2$, $z_3$. Using $z_0$ and give considerations to the equations (1) and (2) above the equations for the foci can be rewritten as (see reference [4], page 680):

$$z_{F_1} = z_0 + \sqrt{z_0^2 - \frac{z_1 * z_2 + z_1 * z_3 + z_2 * z_3}{3}}$$

$$z_{F_2} = z_0 - \sqrt{z_0^2 - \frac{z_1 * z_2 + z_1 * z_3 + z_2 * z_3}{3}}$$

**Box 2 Equations to calculate the foci of the Steiner in-ellipse as functions of the centroid $z_0$ and $z_1, z_2, z_3$**

The Steiner in-ellipse of the triangle $z_1$, $z_2$, $z_3$ is the Steiner circum-ellipse of the triangle $z_{E_1}, z_{E_2}, z_{E_3}$, see reference [8].

## 2.2 Computing the equation of the ellipse[4]

A conic is in general uniquely defined by 5 elements[5] for instance 5 points on it. In appendix (II) it is shown that an ellipse is unambiguously identified if the foci and a point on it are given because in this case one can construct four further points on it.

An ellipse, or more general a conic, in the Cartesian xy or complex plane is represented by an equation of the following form:

$$(I) \quad A * x^2 + B * x * y + C * y^2 + D * x + E * y + F = 0$$

Equation $(I)$ contains the six coefficients $A$ to $F$ although five are needed because one can divide it by any of them which are not zero which leaves five to be calculated[6]. If five points $z_1 = x_1 + i *$ $y_1$ to $z_5 = x_5 + i * y_5$ are given in the complex plane from which not three are collinear then they uniquely define a conic described by equation (I). This means the following system of linear equations holds for any point $z = x + i * y$ which is part of the conic:

$$A * x^2 + B * x * y + C * y^2 + D * x + E * y + F = 0$$

$$A * x_1^2 + B * x_1 * y_1 + C * y_1^2 + D * x_1 + E * y_1 + F = 0$$

$$A * x_2^2 + B * x_2 * y_2 + C * y_2^2 + D * x_2 + E * y_2 + F = 0$$

$$A * x_3^2 + B * x_3 * y_3 + C * y_3^2 + D * x_3 + E * y_3 + F = 0$$

$$A * x_4^2 + B * x_4 * y_4 + C * y_4^2 + D * x_4 + E * y_4 + F = 0$$

$$A * x_5^2 + B * x_5 * y_5 + C * y_5^2 + D * x_5 + E * y_5 + F = 0$$

---

[4] The derivation of the equations for the coefficients A to F given in this section follows
http://www.math.harvard.edu/archive/21b_summer_05/supplements/curves_surfaces.pdf.
[5] For a general discussion of this topic see reference [5], chapter XIV and page 388.
[6] Let $F \neq 0$, then for example (I) can be written as $\frac{A}{F} * x^2 + \frac{B}{F} * x * y + \frac{C}{F} * y^2 + \frac{D}{F} * x + \frac{E}{F} * y + 1 = 0$.

The system of equations can be written in matrix form as:

$$(II) \quad \begin{pmatrix} x^2 & x*y & y^2 & x & y & 1 \\ x_1^2 & x_1*y_1 & y_1^2 & x_1 & y_1 & 1 \\ x_2^2 & x_2*y_2 & y_2^2 & x_2 & y_2 & 1 \\ x_3^2 & x_3*y_3 & y_3^2 & x_3 & y_3 & 1 \\ x_4^2 & x_4*y_4 & y_4^2 & x_4 & y_4 & 1 \\ x_5^2 & x_5*y_5 & y_5^2 & x_5 & y_5 & 1 \end{pmatrix} * \begin{pmatrix} A \\ B \\ C \\ D \\ E \\ F \end{pmatrix} = \begin{pmatrix} 0 \\ 0 \\ 0 \\ 0 \\ 0 \\ 0 \end{pmatrix}$$

Equation $(II)$ has a nontrivial solution only if the determinant of the 6 x 6 square matrix is zero:

$$(III) \quad \begin{vmatrix} x^2 & x*y & y^2 & x & y & 1 \\ x_1^2 & x_1*y_1 & y_1^2 & x_1 & y_1 & 1 \\ x_2^2 & x_2*y_2 & y_2^2 & x_2 & y_2 & 1 \\ x_3^2 & x_3*y_3 & y_3^2 & x_3 & y_3 & 1 \\ x_4^2 & x_4*y_4 & y_4^2 & x_4 & y_4 & 1 \\ x_5^2 & x_5*y_5 & y_5^2 & x_5 & y_5 & 1 \end{vmatrix} = 0$$

If the determinant on the left side of equation $(III)$ is expanded in cofactors along the first row we get:

$$(IV) \; x^2 * \begin{vmatrix} x_1 * y_1 & y_1^2 & x_1 & y_1 & 1 \\ x_2 * y_2 & y_2^2 & x_2 & y_2 & 1 \\ x_3 * y_3 & y_3^2 & x_3 & y_3 & 1 \\ x_4 * y_4 & y_4^2 & x_4 & y_4 & 1 \\ x_5 * y_5 & y_5^2 & x_5 & y_5 & 1 \end{vmatrix} - \begin{vmatrix} x_1^2 & y_1^2 & x_1 & y_1 & 1 \\ x_1^2 & y_2^2 & x_2 & y_2 & 1 \\ x_3^2 & y_3^2 & x_3 & y_3 & 1 \\ x_4^2 & y_4^2 & x_4 & y_4 & 1 \\ x_5^2 & y_5^2 & x_5 & y_5 & 1 \end{vmatrix} * x * y$$

$$+ \begin{vmatrix} x_1^2 & x_1 * y_1 & x_1 & y_1 & 1 \\ x_2^2 & x_2 * y_2 & x_2 & y_2 & 1 \\ x_3^2 & x_3 * y_3 & x_3 & y_3 & 1 \\ x_4^2 & x_4 * y_4 & x_4 & y_4 & 1 \\ x_5^2 & x_5 * y_5 & x_5 & y_5 & 1 \end{vmatrix} * y^2 - \begin{vmatrix} x_1^2 & x_1 * y_1 & y_1^2 & y_1 & 1 \\ x_2^2 & x_2 * y_2 & y_2^2 & y_2 & 1 \\ x_3^2 & x_3 * y_3 & y_3^2 & y_3 & 1 \\ x_4^2 & x_4 * y_4 & y_4^2 & y_4 & 1 \\ x_5^2 & x_5 * y_5 & y_5^2 & y_5 & 1 \end{vmatrix} * x$$

$$+ \begin{vmatrix} x_1^2 & x_1 * y_1 & y_1^2 & x_1 & 1 \\ x_2^2 & x_2 * y_2 & y_2^2 & x_2 & 1 \\ x_3^2 & x_3 * y_3 & y_3^2 & x_3 & 1 \\ x_4^2 & x_4 * y_4 & y_4^2 & x_4 & 1 \\ x_5^2 & x_5 * y_5 & y_5^2 & x_5 & 1 \end{vmatrix} * y - \begin{vmatrix} x_1^2 & x_1 * y_1 & y_1^2 & x_1 & y_1 \\ x_2^2 & x_2 * y_2 & y_2^2 & x_2 & y_2 \\ x_3^2 & x_3 * y_3 & y_3^2 & x_3 & y_3 \\ x_4^2 & x_4 * y_4 & y_4^2 & x_4 & y_4 \\ x_5^2 & x_5 * y_5 & y_5^2 & x_5 & y_5 \end{vmatrix} = 0$$

Comparing $(IV)$ to equation $(I)$ on gets the equations to calculate the coefficients $A$ to $F$ as shown in box 3.

$$A = \begin{vmatrix} x_1 * y_1 & y_1^2 & x_1 & y_1 & 1 \\ x_2 * y_2 & y_2^2 & x_2 & y_2 & 1 \\ x_3 * y_3 & y_3^2 & x_3 & y_3 & 1 \\ x_4 * y_4 & y_4^2 & x_4 & y_4 & 1 \\ x_5 * y_5 & y_5^2 & x_5 & y_5 & 1 \end{vmatrix}$$

$$B = - \begin{vmatrix} x_1^2 & y_1^2 & x_1 & y_1 & 1 \\ x_1^2 & y_2^2 & x_2 & y_2 & 1 \\ x_3^2 & y_3^2 & x_3 & y_3 & 1 \\ x_4^2 & y_4^2 & x_4 & y_4 & 1 \\ x_5^2 & y_5^2 & x_5 & y_5 & 1 \end{vmatrix}$$

$$C = \begin{vmatrix} x_1^2 & x_1 * y_1 & x_1 & y_1 & 1 \\ x_2^2 & x_2 * y_2 & x_2 & y_2 & 1 \\ x_3^2 & x_3 * y_3 & x_3 & y_3 & 1 \\ x_4^2 & x_4 * y_4 & x_4 & y_4 & 1 \\ x_5^2 & x_5 * y_5 & x_5 & y_5 & 1 \end{vmatrix}$$

$$D = - \begin{vmatrix} x_1^2 & x_1 * y_1 & y_1^2 & y_1 & 1 \\ x_2^2 & x_2 * y_2 & y_2^2 & y_2 & 1 \\ x_3^2 & x_3 * y_3 & y_3^2 & y_3 & 1 \\ x_4^2 & x_4 * y_4 & y_4^2 & y_4 & 1 \\ x_5^2 & x_5 * y_5 & y_5^2 & y_5 & 1 \end{vmatrix}$$

$$E = \begin{vmatrix} x_1^2 & x_1 * y_1 & y_1^2 & x_1 & 1 \\ x_2^2 & x_2 * y_2 & y_2^2 & x_2 & 1 \\ x_3^2 & x_3 * y_3 & y_3^2 & x_3 & 1 \\ x_4^2 & x_4 * y_4 & y_4^2 & x_4 & 1 \\ x_5^2 & x_5 * y_5 & y_5^2 & x_5 & 1 \end{vmatrix}$$

$$F = - \begin{vmatrix} x_1^2 & x_1 * y_1 & y_1^2 & x_1 & y_1 \\ x_2^2 & x_2 * y_2 & y_2^2 & x_2 & y_2 \\ x_3^2 & x_3 * y_3 & y_3^2 & x_3 & y_3 \\ x_4^2 & x_4 * y_4 & y_4^2 & x_4 & y_4 \\ x_5^2 & x_5 * y_5 & y_5^2 & x_5 & y_5 \end{vmatrix}$$

**Box 3 Equations to calculate the coefficients $A$ to $F$ of a conic from five given points**

We are now able to estimate the six coefficients $A$ to $F$ in equation $(I)$ given 5 points from which three are not collinear. The next question we have to answer is:

- How can we get at least two more points which are part of the ellipse defined by the foci $z_{F_1}, z_{F_2}$ and three points , $z_{E_1}, z_{E_2}, z_{E_3}$ on it?

Here are two answers:

(1) We can estimate three additional points on the conic by reflection of $z_{E_1}, z_{E_2}$ and $z_{E_3}$ across the line through $z_{F_1}$ and $z_{F_2}$ which is the major axis of the conic. We have to consider the special case that two of the three $z_{E_1}, z_{E_2}, z_{E_3}$ are reflected into each other. This happens if the triangle defined by $z_1, z_2, z_3$ is an isosceles triangle. To avoid treating the special case for isosceles triangle we can use the procedure described in b) below to get three more points.

(2) We can estimate three additional points on the conic by reflections of $z_{E_1}$, $z_{E_2}$ and $z_{E_3}$ in the point $z_0 = \frac{z_{F_1} + z_{F_2}}{2}$ which is the center of the conic. This ensures that the result of the reflections is not in $\{z_{E_1}, z_{E_2}, z_{E_3}\}$[7].

We will choose method (2) to calculate the required additional points, because we do not have to consider the special case of an isosceles triangle.

The reflection of $z$ in $z_0$ is given according to reference [1], page 166, by the equation

$$z_0 = 2 * z_0 - z = \frac{2*z_1 + 2*z_2 + 2*z_3}{3} - z.$$

Putting the values for $z_{E_1}$, $z_{E_2}$ and $z_{E_3}$ into this equation we get the following equations for the reflected points:

$$z_{E_{1r}} = \frac{z_1 + z_2 + 4 * z_3}{6}$$
$$z_{E_{2r}} = \frac{z_1 + 4 * z_2 + z_3}{6}$$
$$z_{E_{3r}} = \frac{4 * z_1 + z_2 + z_3}{6}$$

**Box 4 Equations to calculate 3 additional points on the Steiner in-ellipse by reflecting $z_{E_1}$, $z_{E_2}$ and $z_{E_3}$ in $z_0$**

We decide to take the points $z_{E_1}, z_{E_2}, z_{E_3}, z_{E_{2r}}$ and $z_{E_{3r}}$ to calculate the coefficients $A$ to $F$ using the equations given in box 3.

---

[7] Points of an ellipse are reflections of each other in the center of the ellipse if the tangents in these points are parallel. This cannot happen if the tangent points are on tangents which make up a triangle.

# 3 Scilab implementation

Before we really start with the implementation of the Scilab application to visualize Marden's theorem it is a good idea to write down a brief specification of what it should provide.

## 3.1 Brief Specification of the Scilab application to visualize Marden's theorem

The following steps shall be executed by the application in a loop:

(1) Input of the values $z_1$, $z_2$ and $z_3$

> The application shall provide a form to prompt for the input of the three complex numbers $z_1$, $z_2$, $z_3$ which build the vertices of the triangle for which the Steiner in-ellipse has to be calculated. The form should present default values which can be changed by the user. The form shall have two buttons "*Ok*" and "*Cancel*". A click on the first button triggers the start of the calculations a click on the second terminates the application. The input form shall remember the last input values for $z_1$, $z_2$ and $z_3$ and provide these as default values during the next execution of the loop. If the user clicks on the **"Ok"**-button the input values – the complex numbers $z_1$, $z_2$, $z_3$ – are checked if they are collinear. If they are, a message box with an **"OK"**-button shall inform the user that the input values are collinear. If he clicks the **"Ok"**-Button the input form shows up again and he can change the values of $z_1$, $z_2$, $z_3$ or leave the application by clicking on the "*Cancel*"-button. If the input values for $z_1$, $z_2$ and $z_3$ are not collinear the following steps (2) – (3) are executed.

(2) Calculating the foci $z_{F_1}$ and $z_{F_2}$ the semi major axis and the equation parameters of the Steiner in-ellipse

> The foci $z_{F_1}$ and $z_{F_2}$ of the Steiner in-ellipse are calculated, the three centers $z_{E_1}$, $z_{E_2}$ and $z_{E_3}$ of the sides of the triangle $z_1$, $z_2$ and $z_3$ which are points on the ellipse as well as its center $z_0$ (see Box 1). Additional three points $z_{E_{1r}}$, $z_{E_{2r}}$ and $z_{E_{3r}}$ on the ellipse are calculated by reflecting $z_{E_1}$, $z_{E_2}$ and $z_{E_3}$ in $z_0$ (see box 3). The five points $z_{E_1}$, $z_{E_2}$, $z_{E_3}$, $z_{E_{2r}}$ and $z_{E_{3r}}$ are chosen to compute the coefficients $A$ to $F$ of the equation of the ellipse using the equations in box 3.

(3) Display the triangle and Steiner in-ellipse in a Scilab graphics window

> The triangle $z_1$, $z_2$, $z_3$ its Steiner in-ellipse whose equation is known from step (2) its center $z_0$ as well as its foci $z_{F_1}$ and $z_{F_2}$ are plotted. The graphics also shows the coordinate grid and a title which contains the values of the foci and the values of the semi major and minor axes of the Steiner in-ellipse and its eccentricity. The user is prompted for new values of $z_1$, $z_2$ and $z_3$ which he can enter by the procedure described above and then clicks on the **"Ok"**-button to get a new graphics or he can click the "Cancel"-button to leave the application.

## 3.2 Input of the values $z_1$, $z_2$ and $z_3$ and "main" program

First we have to get the values of the complex numbers $z_1$, $z_2$, $z_3$ from the user. This task is implemented in the Scilab language as shown in listing 1 as a function **inputTrianglePoints($z_1$, $z_2$, $z_3$)**. The function takes three complex numbers as parmeters and shows them in a dialog box to the user as can be seen in figure 1.

```
inputTrianglePoints.sci  ✕
1  function [z1, z2, z3, collinear, loopexit] = inputTrianglePoints(z1, z2, z3)
2      loopexit = %F
3      txt = ['z1';'z2';'z3'];
4      z1s = string(z1)
5      z2s = string(z2)
6      z3s = string(z3)
7      res = x_mdialog('Enter 3 complex numbers', txt, [z1s; z2s; z3s])
8      if length(res) <> 0 then
9          z1 = evstr(res(1))
10         z2 = evstr(res(2))
11         z3 = evstr(res(3))
12         test = imag((z3 - z1) / (z2 - z1))
13         collinear = (test == 0.0)
14         printf("\nimag((z3 - z1) / (z2 - z1)) = %f\n", test)
15     else
16         collinear = %T
17         loopexit = %T
18     end
19 endfunction
```

**Listing 1**

The user is able to change the values for $z_1$, $z_2$ and $z_3$. In listing 1 this form is initialized in line 7 using the Scilab provided function **x_mdialog(…)**. For more details about it see the Scilab documentation.

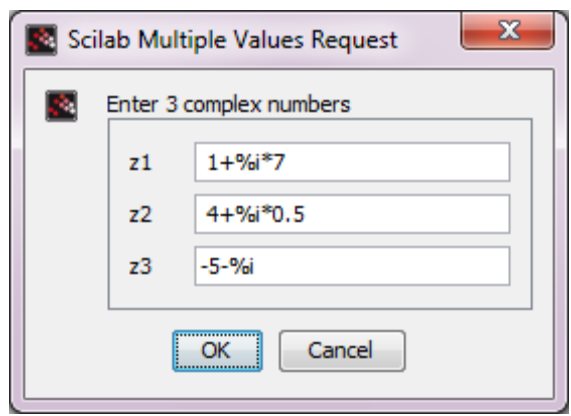

**Figure 1 Input form for the values of $z_1$, $z_2$, $z_3$**

If the user clicks **"Cancel"** the variable **res** in line 7 in listing 1 is an empty matrix and the function will return the Boolean value true (%T in Scilab) in the variable **loopexit** (line 16 in listing 1) which can be used to exit the surrounding loop of the function. If the user clicks **"Ok"** the input from the form is assigned to the complex numbers $z_1$, $z_2$ and $z_3$ in line $9 - 11$ of listing 1. Then in line 12 it is calculated if they are collinear. They are not collinear if the following inequality holds (see reference [1], page 70) $imag\left(\frac{z_3 - z_1}{z_2 - z_1}\right) \neq 0$ ($imag$ gets the imaginary part of a complex number in Scilab). In line 13 the variable **collinear** is assigned the Boolean value true if $z_1$, $z_2$ and $z_3$ are collinear otherwise it gets the value false. The function returns the values of $z_1$, $z_2$, $z_3$, **collinear** and ***loopexit*** in an array. The function ***inputTrianglePoints*** is called in a loop in the "main" program. The "main" program is a Scilab script, stored in a file with the name ***mardensTheorem.sce***. It has a simple structure as can be seen in listing 2 below.

```
mardensTheorem.sce

1  clc();
2  format('v', 5);
3  lib("F:\CASTest\Scilab\MardensTheorem\mardenLib");
4  z1 = complex(1, 7);
5  z2 = complex(4, -0.5);
6  z3 = complex(-5, -1);
7  loopexit = %F;
8  while ~loopexit do
9      [z1, z2, z3, collinear, loopexit] = inputTrianglePoints(z1, z2, z3);
10     if loopexit then
11         printf("\nBye bye!!\n");
12     else
13         if ~collinear then
14             plotSteinerInEllipse(z1, z2, z3);
15         else
16             z1s = string(z1);
17             z2s = string(z2);
18             z3s = string(z3);
19             msg = msprintf("z1=%s, z2=%s z3=%s are collinear!", z1s, z2s, z3s);
20             messagebox(msg, "Collinearity", "info", "modal");
21         end
22     end
23 end
```

**Listing 2 "Main" program to draw a triangle and its Steiner in-ellipse**

In the first line the Scilab command windows is cleared then the format for the string representation of floating-point numbers is set. In line 3 the Scilab ***lib*** function is used to make all functions known to the script which are in the directory ***F:\CASTest\Scilab\MardensTheorem\mardenLib***. This directory contains functions in ***\*.sci*** files as well as the compiled functions in ***\*.bin*** files. This is a Scilab proprietary binary format. The functions stored in this folder have been developed to solve our overall programming task. We will discuss them in the following sections. Now we go back to listing 2. In line $4-6$ the complex variables $z_1$, $z_2$, $z_3$ are assigned initial values which are not collinear. Then the variable ***loopexit*** is set to Boolean value false and the while-loop is entered in line 8. The loop is left if the function ***inputTrianglePoints*** return false in ***loopexit***. If this is the case "Bye bye!!" is printed to the command window of Scilab and the script is terminated. Otherwise if $z_1$, $z_2$, $z_3$ are not collinear the function $plotSteinerInEllipse(z_1, z_2, z_3)$ is called. If they are collinear a message box is shown to the user and the function ***inputTrianglePoints*** is called again. The workhorse of the application is the function $plotSteinerInEllipse$ which is shown in listing 3. Its major task is to plot the triangle and the Steiner in-ellipse into a Scilab graphics window. To achieve this it calls a couple of other functions which are discussed in the following sections. We will start with the function ***computeSteinerInEllipse*** in the next chapter. It is called in line 2 of listing 3 .

```
     plotSteinerInEllipse.sci  ⊠
1    // Copyright (C) Klaus Rohe, D-85625 Glonn
1    function plotSteinerInEllipse(z1, z2, z3)
2        [A, B, C, D, E, F, zF1, zF2,a , b] = computeSteinerInEllipse(z1, z2, z3)
3        // Calculate the center z0 of the Steiner in-ellipse
4        z0 = (zF1 + zF2) / 2
5        printEllipseParameters(A, B, C, D, E, F, a, b, zF1, zF2)
6        [xmin, ymin, xmax, ymax] = computexySquare(z1, z2, z3)
7        initGraphics(xmin, ymin, xmax, ymax)
8        // Plot the traingle defined by z1, z2, z3
9        displayTriangleStuff(z1, z2, z3)
10       deff("res=steinerInEllipseFunc(x, y)","res = A*x^2+B*x*y+C*y^2+D*x+E*y+F")
11       data = linspace(xmin - 3.5, xmax + 3.5, 1000)
12       x = data
13       y = data
14       // Display Steiner in-ellipse
15       contour2d(x, y, steinerInEllipseFunc, [0, 0], style=color("red"), axesflag=1)
16       // Plot z0 the center of the Steiner in-ellipse
17       // as a small blue circle
18       plot([real(z0), real(z0)], [imag(z0), imag(z0)], "bo")
19       xstringb(real(z0), imag(z0), "z0", 0.5, 0.5)
20       // If the eccentricity of the ellipse is big enough display
21       // the foci and the line which connects them
22       if sqrt(a^2 - b^2) / a > 0.1 then
23           // Plot the foci zF1 and zF2 of the Steiner in-ellipse
24           // as small red circles
25           plot([real(zF1), real(zF2)], [imag(zF1), imag(zF2)], "ro")
26           xstringb(real(zF1), imag(zF1), "zF1", 0.5, 0.5)
27           xstringb(real(zF2), imag(zF2), "zF2", 0.5, 0.5)
28           plotLine(zF1, zF2)
29       end
30   endfunction
```

**Listing 3 The function *plotSteinerInEllipse* is the work horse of the Scilab application**

### 3.3 Calculating the foci $z_{F_1}$ and $z_{F_2}$ the semi major axis and the equation parameters of the Steiner in-ellipse

The function $computeSteinerInEllipse(z_1, z_2, z_3)$ shown in listing 4 takes as input parameters the complex numbers $z_1$, $z_2$ and $z_3$ and uses them to calculate the foci $z_{F_1}, z_{F_2}$ of the Steiner in-ellipse its semi major and minor axis and the six coefficients $A$ to $F$ of the equation of the ellipse. It returns them as the array *[A, B, C, D, E, F, a , b, zF1, zF2]*. The function also writes information about intermediate results to the Scilab command window. See lines $2, 5 - 7, 17, 16 - 19$ and line 21 in listing 4.

### 3.3.1 Calculating the foci $z_{F_1}$ and $z_{F_2}$ of the Steiner in-ellipse and three points on it

In line 3 of listing 4 the function $steinerInEllipsePoints(z_1, z_2, z_3)$ is called which uses the equations given in box 1 to calculate the foci and three points on the ellipse. The result is returned in the array *[zE1, zE2, zE3, zF1, zF2]*. The Scilab source code of this function is shown in listing 5. In line 4 of listing 4 the semi major and minor axes of the ellipse are computed by a call to the function $computeAxis(zE, zF1, zF2)$. The parameters of the function are three complex numbers the first is on the ellipse and the second and third are the foci of the ellipse. The semi major and minor axes are returned in the array *[a, b]*. The function computes the value of the semi major axis *a* of the ellipse using the fact that the sum of the distances from any point on the ellipse to its foci is $2 * a$.

The value of the semi major axis *b* is calculated using the formula $b = \sqrt{a^2 - \frac{1}{4} * abs^2(z_{F_1} - z_{F_2})}$, which derives from the fact that the sum of the square of the half distance of the foci and the square of semi minor axis *b* equals the square of the semi major axis *a*, see figure 2.

```
computeSteinerInEllipse.sci
1  //-----------------------------------------------------------------------
2  // Copyright (C) Klaus Rohe, D-85625 Glonn
3  //
4  // Computes the equation of the Steiner in-ellipse of the triangle given by the
5  // complex numbers z1, z2 and z3. It is assumed that z1, z2 and z3 are not
6  // collinear.
7  // The return values are:
8  //      The parameters A, B, C, D, E and F
9  //      for the equation of the ellipse in the form
10 //      A * x^2 + B * x * y + C * y^2 + D * x + E * y + F = 0
11 //      The foci zF1 and zF2 of the ellipse.
12 //      The length a and b of the semi major and minor axes of the ellipse.
13 //-----------------------------------------------------------------------
1  function [A, B, C, D, E, F, zF1, zF2, a, b] = computeSteinerInEllipse(z1, z2, z3)
2      printf("\nEntering function computeSteinerInEllipse\n")
3      [zF1, zF2, zE1, zE2, zE3] = steinerInEllipsePoints(z1, z2, z3)
4      [a, b] = computeAxis(zE1, zF1, zF2)
5      printf("\tzE1=%s\n", string(zE1))
6      printf("\tzE2=%s\n", string(zE2))
7      printf("\tzE3=%s\n", string(zE3))
8      // Midpoint of the ellipse
9      z0 = (zF1 + zF2) / 2
10     // Reflect at the center (zF1 + zF2) / 2 of the Steiner ellipse.
11     zE1r = (z1 + z2 + 4 * z3) / 6
12     zE2r = (z1 + 4 * z2 + z3) / 6
13     zE3r = (4 * z1 + z2 + z3) / 6
14     printf("\tReflection in center z0 = (zF1 + zF2) / 2 =%s\n", string(z0))
15     printf("\tzE1r=%s\n", string(zE1r))
16     printf("\tzE2r=%s\n", string(zE2r))
17     printf("\tzE3r=%s\n", string(zE3r))
18     [A, B, C, D, E, F] = conicByFivePoints(zE1, zE2, zE3, zE2r, zE3r)
19     printf("Leaving function computeSteinerInEllipse\n\n")
20 endfunction
```

**Listing 4 Scilab source code of the function *computeSteinerInEllipse***

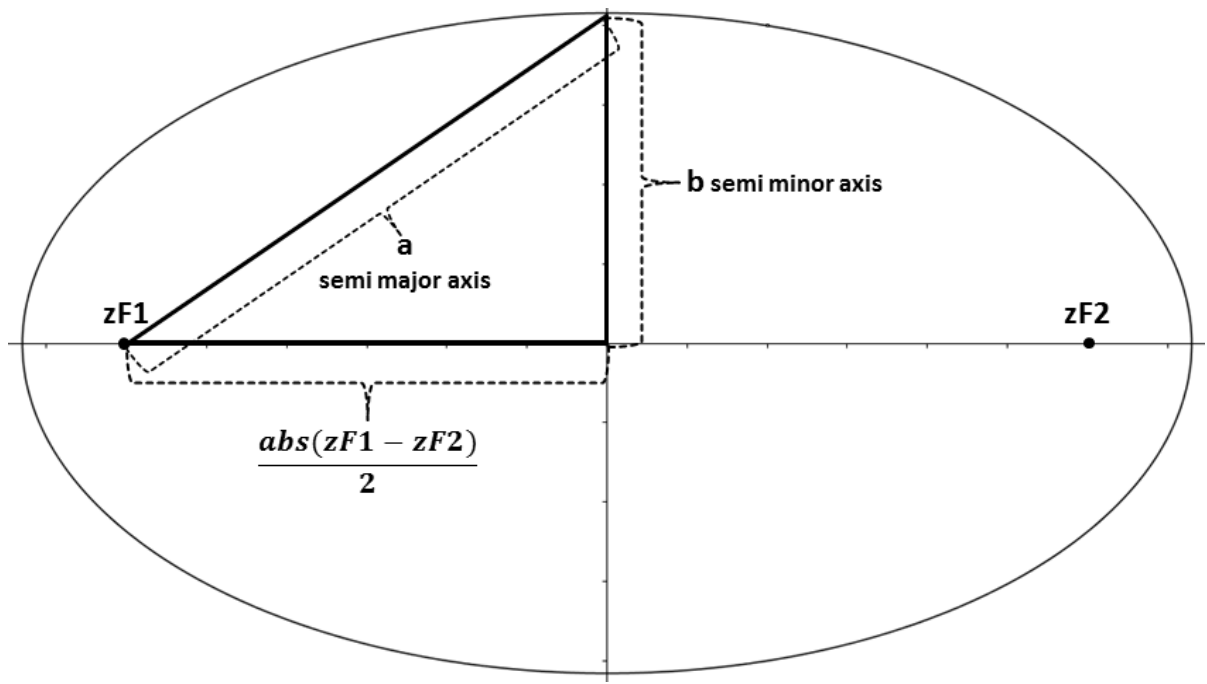

**Figure 2** Relation between semi major axis and semi minor axis $b^2 + \left[\frac{abs(zF1 - zF2)}{2}\right]^2 = a^2$ (theorem of Pythagoras), $abs(zF1 - zF2)$ is the distance of the foci zF1 and zF2.

```
steinerInEllipsePoints.sci  ⊠

1  function [zF1, zF2, zE1, zE2, zE3] = steinerInEllipsePoints(z1, z2, z3)
2      zE1 = (z1 + z2) / 2
3      zE2 = (z1 + z3) / 2
4      zE3 = (z2 + z3) / 2
5      D = sqrt(z1^2 + z2^2 + z3^2 -(z1 * z2 + z1 * z3 + z2 *z3))
6      zF1 = (z1 + z2 + z3 + D) / 3
7      zF2 = (z1 + z2 + z3 - D) / 3
8  endfunction
```

**Listing 5 Calculating the foci and 3 points on the Steiner in-ellipse**

The Scilab source code of function **computeAxis** is shown in listing 6. It makes use of the above formula for the calculation of the semi minor axis of an ellipse.

```
computeAxis.sci  ⊠

1  function [a, b] = computeAxis(zE, zF1, zF2)
2      // Semi major axis of the ellipse
3      a = 0.5 * (abs(zE - zF1) + abs(zE - zF2))
4      // Semi minor axis of the ellipse
5      b = sqrt(a^2 - 0.25 * abs(zF1 - zF2)^2)
6  endfunction
```

**Listing 6  Computing the semi major and minor axis given a point on the ellipse and its foci**

### 3.3.2 Calculating additional points on the ellipse to determine its equation

The function $steinerInEllipsePoints(z1, z2, z3)$ returns among others the three points **zE1, zE2** and **zE3** which are on the ellipse. To compute the equation of the ellipse we need five points as discussed in section 2.2. We can compute three additional points on the ellipse by reflecting **zE1, zE2** and **zE3** in its center **z0**. This is done in the function **computeSteinerInEllipse** using the equations in box 4. See listing 4 lines 11 – 13. Now we have three additional points **zE1r, zE2r** and **zE3r** on the ellipse. We choose **zE1, zE2, zE3, zE2r** and **zE3r** to compute the six coefficients **A, B, C, D, E** and **F** of the equation of the ellipse using the formulas shown in box 2. This is implemented in the function $conicByFivePoints(z1, z2, z3, z4, z5)$ which takes five complex numbers which are part of the conic and returns the coefficients of the equation of the conic in the array **[A, B, C, D, E, F]**. The Scilab source code of this function is shown in listing 7.It uses the Scilab provided function **det** to compute the determinant of a square matrix.

conicByFivePoints.sci

```
 1  //-----------------------------------------------------------
 2  // Computes the parameters A, B, C, D and E of the conic
 3  // A * x^2 + B * x * y + C * y^2 + D * x + E * y + F = 0
 4  // which passes through the five points represented by the
 5  // complex numbers z1, z2, z3, z4 and z5.
 6  //-----------------------------------------------------------
 1  function [A, B, C, D, E, F] = conicByFivePoints(z1, z2, z3, z4, z5)
 2      x1 = real(z1)
 3      y1 = imag(z1)
 4      x2 = real(z2)
 5      y2 = imag(z2)
 6      x3 = real(z3)
 7      y3 = imag(z3)
 8      x4 = real(z4)
 9      y4 = imag(z4)
10      x5 = real(z5)
11      y5 = imag(z5)
12      A = det([x1 * y1, y1^2, x1, y1, 1;
13               x2 * y2, y2^2, x2, y2, 1;
14               x3 * y3, y3^2, x3, y3, 1;
15               x4 * y4, y4^2, x4, y4, 1;
16               x5 * y5, y5^2, x5, y5, 1])
17      B = -det([x1^2, y1^2, x1, y1, 1;
18                x2^2, y2^2, x2, y2, 1;
19                x3^2, y3^2, x3, y3, 1;
20                x4^2, y4^2, x4, y4, 1;
21                x5^2, y5^2, x5, y5, 1])
22      C = det([x1^2, x1 * y1, x1, y1, 1;
23               x2^2, x2 * y2, x2, y2, 1;
24               x3^2, x3 * y3, x3, y3, 1;
25               x4^2, x4 * y4, x4, y4, 1;
26               x5^2, x5 * y5, x5, y5, 1])
27      D = -det([x1^2, x1 * y1, y1^2, y1, 1;
28                x2^2, x2 * y2, y2^2, y2, 1;
29                x3^2, x3 * y3, y3^2, y3, 1;
30                x4^2, x4 * y4, y4^2, y4, 1;
31                x5^2, x5 * y5, y5^2, y5, 1])
32      E = det([x1^2, x1 * y1, y1^2, x1, 1;
33               x2^2, x2 * y2, y2^2, x2, 1;
34               x3^2, x3 * y3, y3^2, x3, 1;
35               x4^2, x4 * y4, y4^2, x4, 1;
36               x5^2, x5 * y5, y5^2, x5, 1])
37      F = -det([x1^2, x1 * y1, y1^2, x1, y1;
38                x2^2, x2 * y2, y2^2, x2, y2;
39                x3^2, x3 * y3, y3^2, x3, y3;
40                x4^2, x4 * y4, y4^2, x4, y4;
41                x5^2, x5 * y5, y5^2, x5, y5])
42  endfunction
```

**Listing 7 Scilab implementation of the procedure to compute the *A, B, C, D, E* to *F* of the equation of a conic defined by five complex numbers z1 to z5 as described in section 2.2**

## 3.4 Display the triangle and its Steiner in-ellipse in a graphics window

Now we return to the source code of the function **plotSteinerInEllipse** shown in listing 3. In line 5 of listing 3 the parameters of the Steiner in-ellipse are printed to the Scilab command window by the function **printEllipseParameters**. The Scilab source of this function is shown in listing 8.

```
printEllipseParameters.sci

1  function printEllipseParameters(A, B, C, D, E, F, a, b, zF1, ZF2)
2      mprintf("Parameters of the ellipse\n")
3      mprintf("\tA = %+3.2f\n", A)
4      mprintf("\tB = %+3.2f\n", B)
5      mprintf("\tC = %+3.2f\n", C)
6      mprintf("\tD = %+3.2f\n", D)
7      mprintf("\tE = %+3.2f\n", E)
8      mprintf("\tF = %+3.2f\n", F)
9      mprintf("\tFoci              zF1: %s\n", string(zF1))
10     mprintf("\t                  zF2: %s\n", string(ZF2))
11     mprintf("\tSemi major axis a: %+3.2f\n", a)
12     mprintf("\tSemi minor axis b: %+3.2f\n", b)
13     // Eccentricity of the ellipse
14     e = sqrt(a^2 - b^2) / a
15     mprintf("\tEccentricity eps: %+3.2f\n", e)
16 endfunction
```

**Listing 8 Formatted printing of the ellipse parameters to the Scilab command window**

Then the lower left and upper right coordinates of the vertices of a square are computed which is of sufficient size to allow the drawing of the full triangle $z_1$, $z_2$, $z_3$ in it. This is done by the function **computexySquare** in line 6 of listing 3. The source of this function is in listing 9 below.

```
computexySquare.sci

1  function [xmin, ymin, xmax, ymax] = computexySquare(z1, z2, z3)
2      z0 = (z1 + z2 + z3) / 3 // Centroid of triangle z1, z2, z3
3      x1 = abs(real(z0) - real(z1))
4      x2 = abs(real(z0) - real(z2))
5      x3 = abs(real(z0) - real(z3))
6      y1 = abs(imag(z0) - imag(z1))
7      y2 = abs(imag(z0) - imag(z2))
8      y3 = abs(imag(z0) - imag(z3))
9      dmax = max(x1, x2, x3, y1, y2, y3) + 0.5
10     xmin = real(z0) - dmax
11     ymin = imag(z0) - dmax
12     xmax = real(z0) + dmax
13     ymax = imag(z0) + dmax
14 endfunction
```

**Listing 9**

In line 7 of listing 3 the graphics window is initialized by calling the function **initGraphics**.

```
initGraphics.sci

1  function initGraphics(xmin, ymin, xmax, ymax)
2      clf()
3      xset("wdim", 900, 900)
4      xset("fpf", ".")
5      xset("thickness", 1)
6      xset("font", 10, 3.5)
7      xgrid(2)
8      square(xmin, ymin, xmax, ymax)
9  endfunction
```

**Listing 10**

As can be seen in listing 10 it uses some Scilab provided functions to set initial properties of the graphics window. For the description of these functions see the Scilab documentation. After the initialization of the graphics windows of Scilab the triangle $z_1$, $z_2$, $z_3$ is plotted into the graphics window. This is done by calling the function $\textbf{\textit{displayTriangleStuff}}(z_1, z_2, z_3)$ in line 9 in listing 3. The source of this function is displayed in listing 11. It calls the Scilab provided function $\textbf{\textit{title}}$ which expects

```scilab
displayTriangleStuff.sci
1  function displayTriangleStuff(z1, z2, z3)
2      title(buildTitleString(z1, z2, z3), "fontsize", 3)
3      // Display triangle
4      plotTriangle(z1, z2, z3)
5      xstringb(real(z1), imag(z1), "z1", 0.5, 0.5)
6      xstringb(real(z2), imag(z2), "z2", 0.5, 0.5)
7      xstringb(real(z3), imag(z3), "z3", 0.5, 0.5)
8  endfunction
```

**Listing 11**

a string as a parameter and displays it in the graphics window. The string itself is generated by the function $\textbf{\textit{buildTitleString}}$. The source of this function is shown in listing 12.

```scilab
buildTitleString.sci
1  function titlestr = buildTitleString(z1, z2, z3)
2      [zF1, zF2, _, _, _] = steinerInEllipsePoints(z1, z2, z3)
3      [a, b] = computeAxis((z1 + z2) / 2, zF1, zF2)
4      z0 = (z1 + z2 + z3) /3
5      szF1 = string(zF1)
6      szF2 = string(zF2)
7      sz0 = string(z0)
8      s1 = msprintf("zF1=%s, zF2=%s, z0=%s", szF1, szF2, sz0)
9      s2 = msprintf("a=%3.2f, b=%3.2f, e=%3.2f.", a, b, sqrt(a^2 - b^2) / a)
10     tstr = strcat([s1, ' ', s2])
11     titlestr = msprintf("Steiner in-ellipse of triange z1, z2, z3, %s", tstr)
12 endfunction
```

**Listing 12**

```scilab
plotTriangle.sci
1  function plotTriangle(z1, z2, z3)
2      xp = [real(z1), real(z2), real(z3)]
3      yp = [imag(z1), imag(z2), imag(z3)]
4      xpoly(xp, yp, "lines", closed=1)
5  endfunction
```

**Listing 13**

In line 9 of listing 4 $\textbf{\textit{displayTriangleStuff}}$ plots the triangle $z_1$, $z_2$, $z_3$ to the graphics window by calling the function $\textbf{\textit{plotTriangle}}(z_1, z_2, z_3)$. The implementation is presented in listing 13. This function makes use of the Scilab provided $\textbf{\textit{xpoly}}$ function which can used to plot arbitrary polygons.

Now we are looking at line 10 in listing 3 where the Scilab provided function $\textbf{\textit{deff}}$ is used to define the function $\textbf{\textit{steinerInEllipseFunc}}$ on the fly. The function takes the two arguments $x, y$ and returns the value of the expression $A * x^2 + B * x * y + C * y^2 + D * x + E * y + F$. The values of the coefficients $A$ to $F$ are provided by the call to $\textbf{\textit{computeSteinerInEllipse}}$ in line 2 of listing 3.

The function **steinerInEllipseFunc** will be evaluated in the complex plane on a square defined in line $8-10$ in listing 3. This is done by the call to **contour2d** in line 15. It is a Scilab function which supports the plot of level curves of a surface defined by a real valued function $z = f(x, y)$. For details see the Scilab documentation. In line 18 the **plotd** function of Scilab is applied to plot the center of the Steiner in-ellipse as a small blue circle and in the following line it is marked with the label **z0** using Scilabs **xstringb** function. Line 22 contains a test whether the eccentricity of the Steiner ellipse is bigger than 0.1. If this is the foci of the ellipse are plotted as small red circles in line 25 and are labeled **zF1** and **zF2** in line $26-27$. In line 28 the foci are connected by a line using the function **plotLine($z_1, z_2$)**. Its source code can be seen in listing 14.

```
1 function plotLine(z1, z2)
2     xp = [real(z1), real(z2)]
3     yp = [imag(z1), imag(z2)]
4     xpoly(xp, yp, "lines", closed=0)
5 endfunction
```

**Listing 14**

We have now reached the end of the discussion of the Scilab implementation of the application to visualize of Marden's theorem. In the following chapter it is applied to the three cases: a general, an isosceles and an equilateral triangle and it is presented how the results looks like.

# 5 Executing the Scilab application

The application can be started from the Scilab command window using
$exec('mardensTheorem.sce')$; the input dialog shown in figure 1 will show up and we have to
enter the values for $z_1$, $z_2$, $z_3$ or accept the default values. We will now try it for a general triangle
an isosceles and equilateral triangle.

## 5.1 General triangle

If we enter for example $z1 = 3 + \%i * 14$, $z2 = 8.5 - \%i * 1.5$ and $z3 = -6 - \%i * 2$ which
define a general triangle in the complex plane, the application produces the graphics shown in
figure 3.

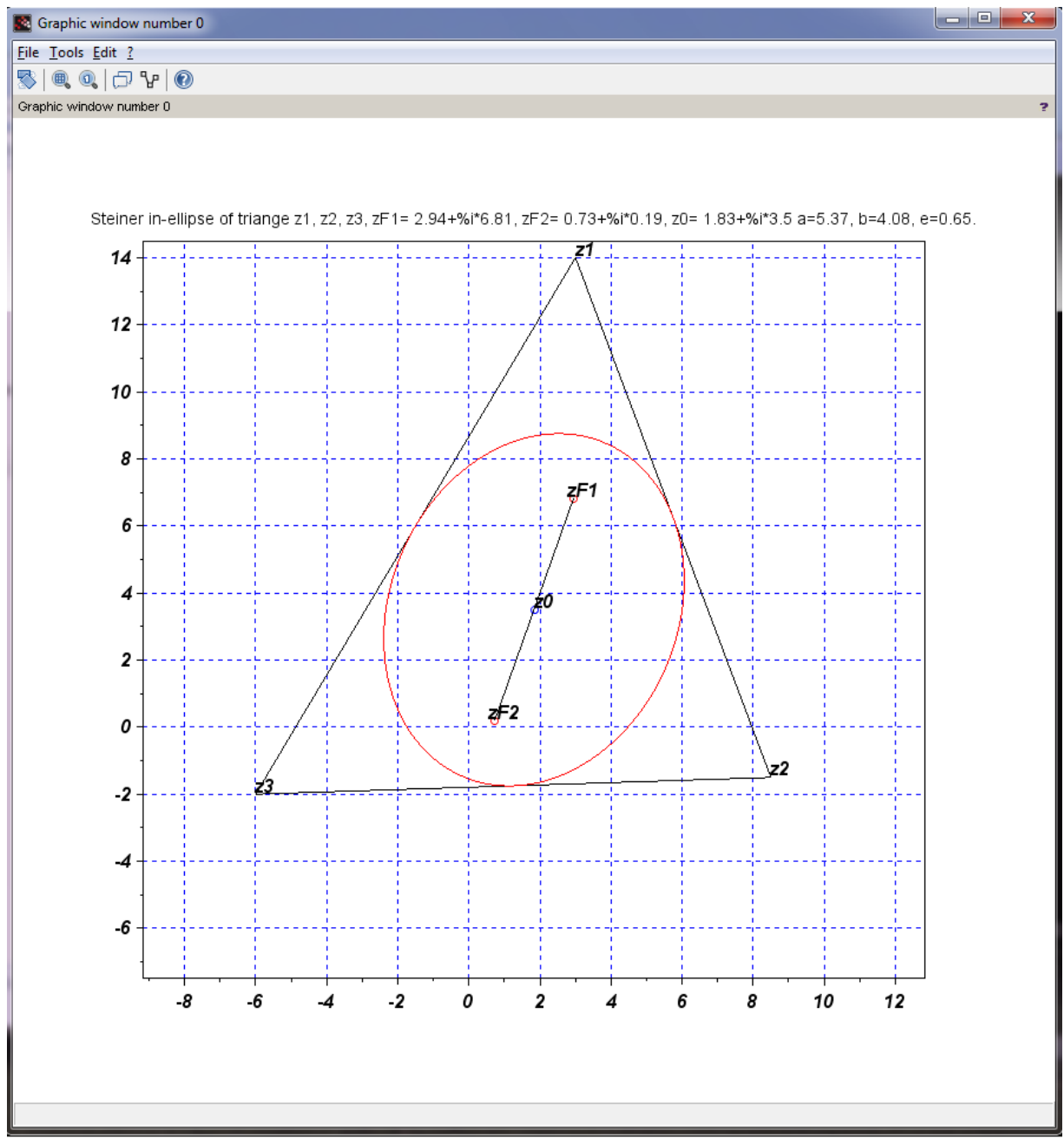

**Figure 3**

## 5.2 Isosceles triangle

If we enter for example $z1 = \%i * 15$, $z2 = 8 - \%i * 2$ and $z3 = -8 - \%i * 2$ which define an isosceles triangle in the complex plane, the application generates the graphics shown in figure 4.

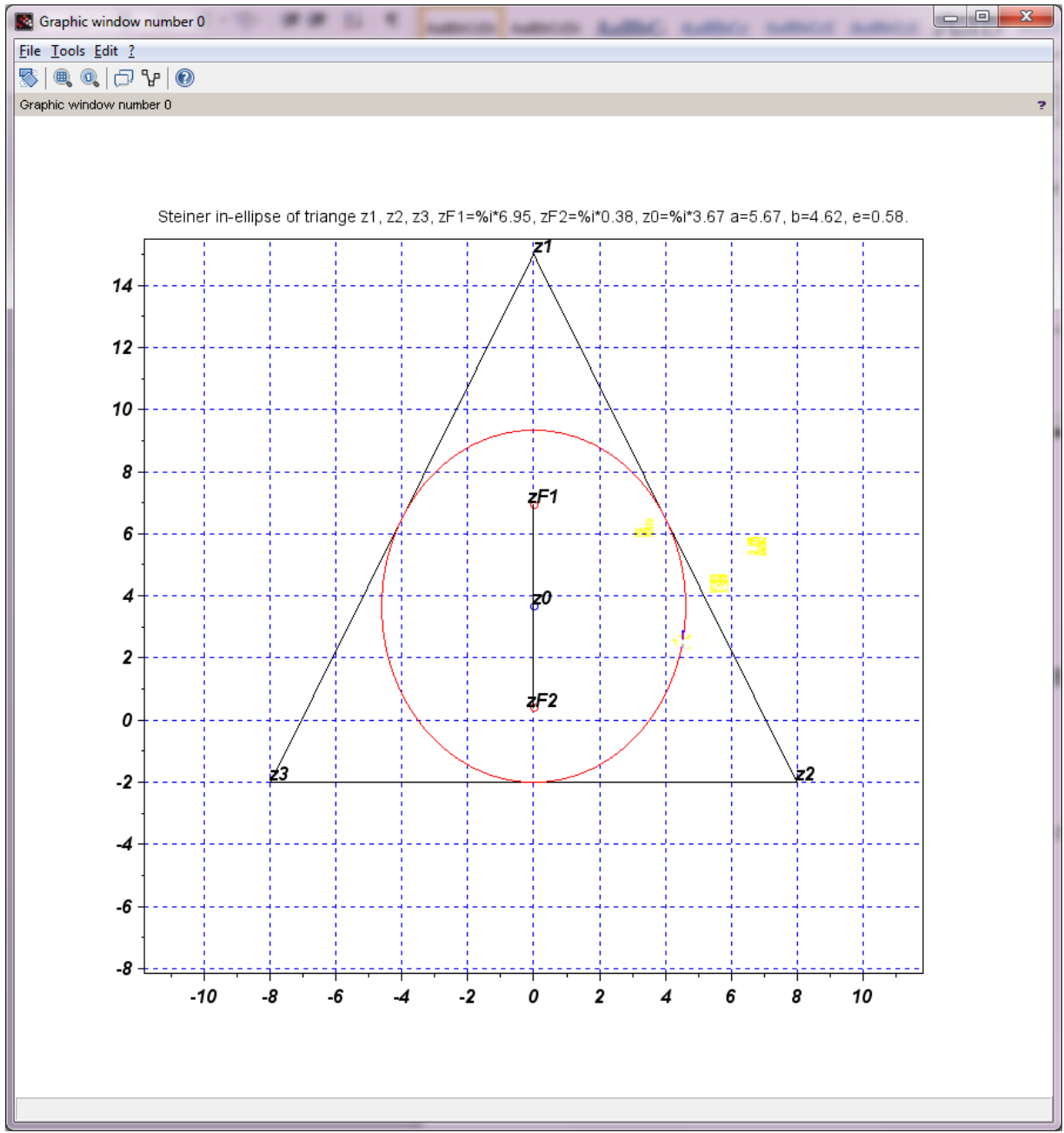

**Figure 4**

## 5.3 Equilateral triangle

We try another example with $z1 = 4$, $z2 = -2 + \%i * sqrt(3) * 2$ and $z3 = -2 - \%i * sqrt(3) * 2$. In this case $z_1$, $z_2$ and $z_3$ are the vertices of an equilateral triangle and the application generates the graphics presented in figure 4.

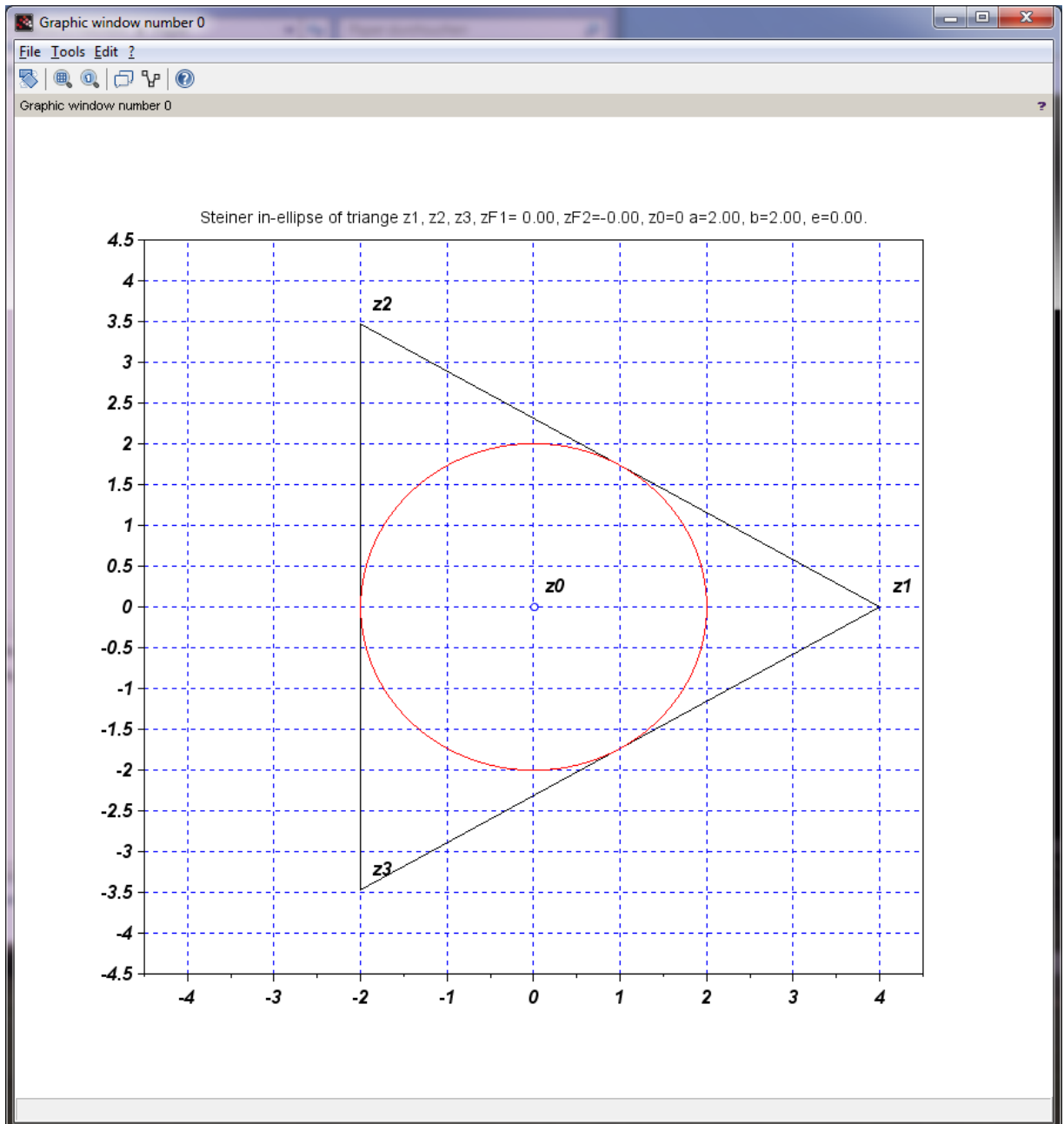

**Figure 5**

One can see that for an equilateral triangle the Steiner in-ellipse and the in-circle are the same because the semi minor and major axis both have the value 2 and the eccentricity is 0.

## Appendix

## (I) The foci and a given point on the ellipse are defining it uniquely

The ellipse is defined by the feature that for every point on the ellipse the sum of it distances to the foci is two times the length of its semi major axis. Let **zF1** and **zF2** the foci of an ellipse in the complex plane and **zE** a point on the ellipse. We will show that from the foci **zF1** and **zF2** and **zE** we can construct four other points on that ellipse.

Let $a$ be the length of the semi major axis. The above feature of the ellipse can then be stated by the following equation $2 * a = abs(zE - zF1) + abs(zE - zF2)$. The length $b$ of the semi minor axis can be computed as shown in section 3.3.

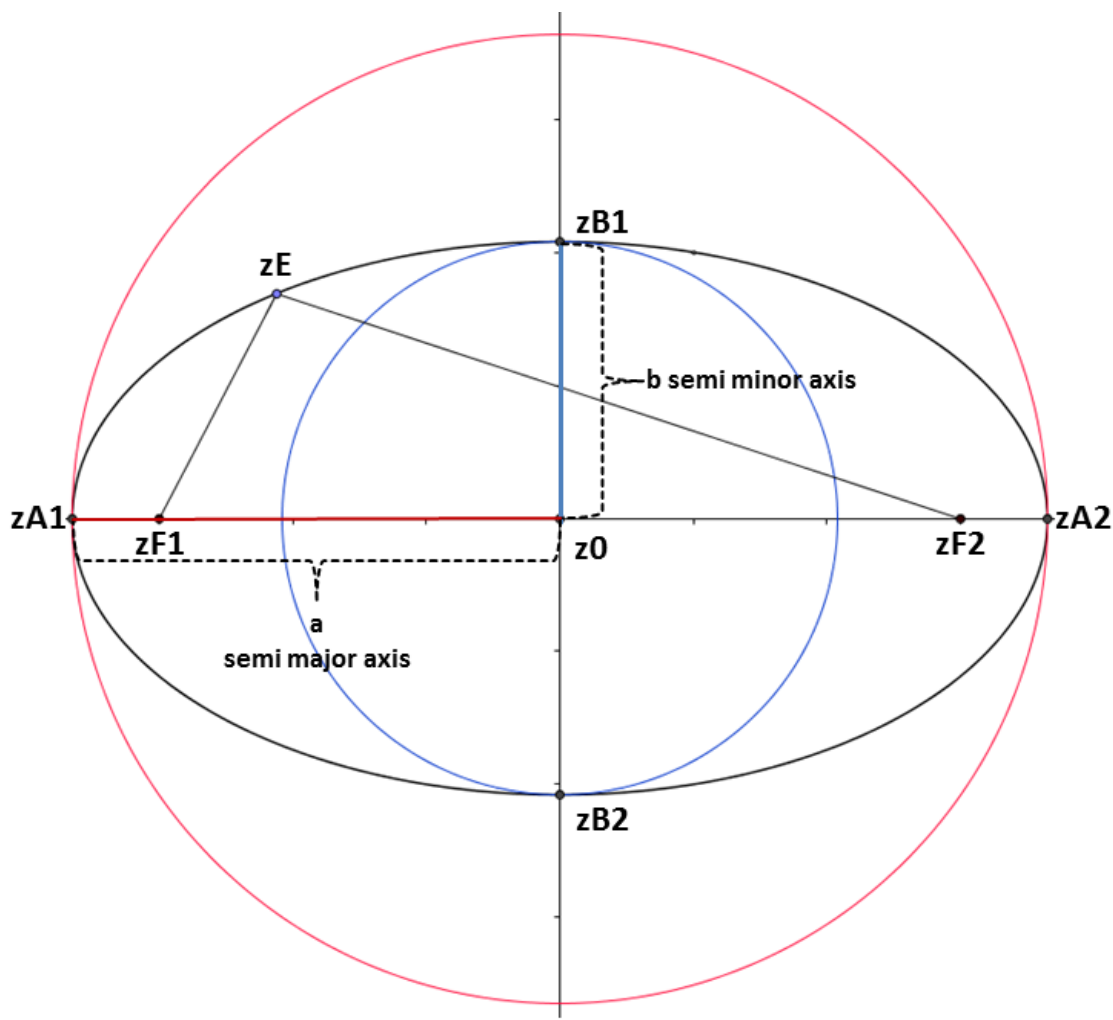

**Figure 6 zF1 and zF2 are the foci of the ellipse z0 its center**

The center of the ellipse is $z0 = \frac{zF1 + FF2}{2}$. Now we draw a circle with radius $a$ around $z0$. It intersects the line containing the semi major axis in the points **zA1** and **zA2** which are points on the ellipse. We draw a second circle with radius $b$ around $z0$. It intersects the line containing the semi minor axis in the points **zB1** and **zB2** which are points on the ellipse too, see figure 3. Now we have the five points **zE, zA1, zA2, zB1** and **zB2** which are on the ellipse and these five points uniquely define it.

## (II) Plotting the Steiner in- and circum-ellipse of a triangle

In reference [2], page 155 there is the following proposition about the Steiner in- and circum-ellipse

> **The conic tangent to the sides of a triangle ABC at their Midpoints is an ellipse $\mathcal{E}$ called the inscribed Steiner ellipse of the triangle; it has the centroid G of ABC for center. The homothety with center G and coefficient -2 transforms the inscribed Steiner ellipse into the circumscribed Steiner ellipse.**

The last part of this proposition gives us a very simple procedure to calculate a point on the Steiner circum-ellipse (*circumscribed Steiner ellipse*) from a given point on the Steiner in-ellipse (*inscribed Steiner ellipse*). According to reference [2], page 28, a homothety ($z \rightarrow z'$) with center $a$ (complex number) and coefficient k (real number) is described by the equation $z' = k * z + a * (1 - k)$. Let $A$ be represented by the complex number $z_1$, $B$ by $z_2$, $C$ by $z_3$ and $G$ by $z_0 = \frac{z_1 + z_2 + z_3}{3}$. For the homothety with center by $z_0$ and $k = -2$ this yields the equation

$z' = -2 * z + z_0 * 3$.

It maps each point on the Steiner in-ellipse to a corresponding point on the Steiner circum-ellipse. From the five points on the Steiner in-ellipse which can be calculated by the procedure outlined in chapter 2.2 we can get five points on the Steiner circum-ellipse. These five points can be used to compute the coefficients of the equation of that ellipse by the algorithm shown in box 3 on page 6.

The above transformation equation is also applied to the foci of the Steiner in-ellipse $zF1_i$, $zF2_{ic}$ to calculate the foci $zF1_c$ and $zF2_c$ of the Steiner circum-ellipse. The result is shown in box 5.

$$zF1_c = z_0 - 2 * \sqrt{z_0{}^2 - \frac{z_1 * z_2 + z_1 * z_3 + z_2 * z_3}{3}}$$

$$zF2_c = z_0 + 2 * \sqrt{z_0{}^2 - \frac{z_1 * z_2 + z_1 * z_3 + z_2 * z_3}{3}}$$

**Box 5 The equations for the foci of the Steiner circum-ellipse as functions of the centroid $z_0$ and $z_1, z_2, z_3$**

All this is implemented in the function **computeSteinerEllipses**$(z_1,\ z_2,\ z_3)$. The Scilab source code is presented in listing 15. It returns the coefficients of the in- and circum-ellipse equations, its foci and the values of the semi major and minor axes in a 2 x 10 matrix of the following structure

$$\begin{pmatrix} A_i & B_i & C_i & D_i & E_i & F_i & zF1_i & zF2_i & a_i & b_i \\ A_c & B_c & C_c & D_c & E_c & F_c & zF1_c & zF2_c & a_c & b_c \end{pmatrix}.$$

The first row of this matrix contains values for the in-ellipse, the second row those for the circum-ellipse. In line $4 - 13$ in listing 15 the foci, the semi major and minor axes and the coefficients of the in-ellipse are calculated by the same procedure as in the function **computeSteinerInEllipse** (see listing 4 on page 14). In line $17 - 18$ the two points $zEc4$ and $zEc5$ are calculated which are on the Steiner circum-ellipse using two points which are on the Steiner in-ellipse by applying the above stated homothety. The points $z_1,\ z_2,\ z_3$ are by definition on the circum-ellipse thus we now have the five points $z_1,\ z_2,\ z_3,\ zEc4,\ zEc5$ which are used to compute the coefficients for the equation of the circum-ellipse by applying the function **conic ByFivePoints**. This is done in line 20 in listing 15. In line 22 and 23 the foci of the circum-ellipse are estimated using the equations in box 5 and in line 24 its semi major and minor axes. In line 25 − 34 the matrix which the function returns as a result, is set up.

```scilab
computeSteinerEllipses.sci

1  //----------------------------------------------------------------------
2  // Copyright (C) Klaus Rohe, D-85625 Glonn
3  //
4  // Computes the equation of the Steiner in- and circum-ellipse of the
5  // triangle given by the complex numbers z1, z2 and z3.
6  // It is assumed that z1, z2 and z3 are not collinear.
7  // The function returns a 2 x 10 Matrix, every row contain the following values:
8  //      The parameters A, B, C, D, E and F for the equation of the ellipses
9  //      in the form A * x^2 + B * x * y + C * y^2 + D * x + E * y + F = 0
10 //      The Foci of the Steiner ellipses.
11 //      The semi major and minor axes of the Steiner ellipses.
12 // The first row of the Matrix contains the values for the Steiner in-ellipse
13 // the second row contains the values for the Steiner circum-ellipse:
14 // [Ai, Bi, Ci, Di, Ei, Fi, ZF1i, ZF2i, ai, bi]
15 // [Ac, Bc, Cc, Dc, Ec, Fc, ZF1c, ZF2c, ac, bc]
16 //----------------------------------------------------------------------
1  function resMatrix = computeSteinerEllipses(z1, z2, z3)
2      resMatrix = zeros(2, 10)
3      printf("\nEntering function computeSteinerEllipses\n")
4      [zF1i, zF2i, zE1, zE2, zE3] = steinerInEllipsePoints(z1, z2, z3)
5      [ai, bi] = computeAxis(zE1, zF1i, zF2i)
6      // Midpoint of the ellipse
7      z0 = (zF1i + zF2i) / 2
8      // Reflect at the center z0 of the Steiner ellipses.
9      zE1r = (z1 + z2 + 4 * z3) / 6
10     zE2r = (z1 + 4 * z2 + z3) / 6
11     zE3r = (4 * z1 + z2 + z3) / 6
12     // Compute coefficients of the in-ellipse equation
13     [Ai, Bi, Ci, Di, Ei, Fi] = conicByFivePoints(zE1, zE2, zE3, zE2r, zE3r)
14     // z1, z2, z3 are already on the Steiner circum-ellipse.
15     // 2 addtional points are calculated applying the homothety
16     // z = -2 * z + z0 * 3 to the points zE1r and zE2r.
17     zEc4 = -2 * zE1r + z0 * 3
18     zEc5 = -2 * zE2r + z0 * 3
19     // Compute coefficients of the circum-ellipse equation
20     [Ac, Bc, Cc, Dc, Ec, Fc] = conicByFivePoints(z1, z2, z3, zEc4, zEc5)
21     // Compute the foci of the Steiner circum-ellipse
22     zF1c = z0 - 2 * sqrt(z0^2 - (z1 * z2 + z1 * z3 + z2 * z3) / 3)
23     zF2c = z0 + 2 * sqrt(z0^2 - (z1 * z2 + z1 * z3 + z2 * z3) / 3)
24     [ac, bc] = computeAxis(zEc4, zF1c, zF2c)
25     resMatrix(1, 1) = Ai; resMatrix(1, 2) = Bi
26     resMatrix(1, 3) = Ci; resMatrix(1, 4) = Di
27     resMatrix(1, 5) = Ei; resMatrix(1, 6) = Fi
28     resMatrix(1, 7) = zF1i; resMatrix(1, 8) = zF2i
29     resMatrix(1, 9) = ai; resMatrix(1, 10) = bi
30     resMatrix(2, 1) = Ac; resMatrix(2, 2) = Bc
31     resMatrix(2, 3) = Cc; resMatrix(2, 4) = Dc
32     resMatrix(2, 5) = Ec; resMatrix(2, 6) = Fc
33     resMatrix(2, 7) = zF1c; resMatrix(2, 8) = zF2c
34     resMatrix(2, 9) = ac; resMatrix(2, 10) = bc
35     printf("Leaving function computeSteinerEllipses\n\n")
36 endfunction
```

**Listing 15 Calculating the equation coefficients, foci and semi major and minor axes of the Steiner ellipses of the triangle** $z_1$, $z_2$, $z_3$

The plotting of the triangle $z_1$, $z_2$, $z_3$ and its Steiner ellipses is done by the function ***plotSteinerEllipses*** whose source code is shown in listing 16.

```
1  // Copyright (C) Klaus Rohe, D-85625 Glonn
1  function plotSteinerEllipses(z1, z2, z3)
2      resMatrix = computeSteinerEllipses(z1, z2, z3)
3      Ai  =  resMatrix(1, 1); Bi  =  resMatrix(1, 2)
4      Ci  =  resMatrix(1, 3); Di  =  resMatrix(1, 4)
5      Ei  =  resMatrix(1, 5); Fi  =  resMatrix(1, 6)
6      zF1i = resMatrix(1, 7); zF2i = resMatrix(1, 8)
7      ai  =  resMatrix(1, 9); bi  =  resMatrix(1, 10)
8      Ac  =  resMatrix(2, 1); Bc  =  resMatrix(2, 2)
9      Cc  =  resMatrix(2, 3); Dc  =  resMatrix(2, 4)
10     Ec  =  resMatrix(2, 5); Fc  =  resMatrix(2, 6)
11     zF1c = resMatrix(2, 7); zF2c = resMatrix(2, 8)
12     ac  =  resMatrix(2, 9); bc  =  resMatrix(2, 10)
13     ei  =  sqrt(1- (bi / ai)^2)
14     ec  =  sqrt(1- (bc / ac)^2)
15     // Print values of the semi major and minor axes and the eccentricities
16     // of the Steiner ellipses to the Scilab command window
17     mprintf("Steiner in-ellipse\n")
18     mprintf("\tValue of semi major axis a = %+3.2f\n", ai)
19     mprintf("\tValue of semi minor axis b = %+3.2f\n", bi)
20     mprintf("\tLinear eccentricity       e = %+3.2f\n", ei)
21     mprintf("Steiner circum-ellipse\n")
22     mprintf("\tValue of semi major axis a = %+3.2f\n", ac)
23     mprintf("\tValue of semi major axis a = %+3.2f\n", bc)
24     mprintf("\tLinear eccentricity       e = %+3.2f\n", ec)
25     // Calculate the center z0 of the Steiner ellipses
26     z0 = (z1 + z2 + z3) / 3
27     [xmin, ymin, xmax, ymax] = computexySquare(z1, z2, z3)
28     initGraphics(xmin, ymin, xmax, ymax)
29     // Plot the traingle defined by z1, z2, z3
30     titleStr =  "Steiner in-ellipse red (foci zF1i, zF2i) and"
31     titleStr = titleStr + "Steiner circum-ellipse blue (foci zF1c, zF2c)"
32     title(titleStr, "fontsize", 3)
33     // Display triangle
34     plotTriangle(z1, z2, z3)
35     xstringb(real(z1), imag(z1), "z1", 0.5, 0.5)
36     xstringb(real(z2), imag(z2), "z2", 0.5, 0.5)
37     xstringb(real(z3), imag(z3), "z3", 0.5, 0.5)
38     // Plot z0 the center of the Steiner ellipsess
39     // as a small blue circle
40     plot([real(z0), real(z0)], [imag(z0), imag(z0)], "go")
41     xstringb(real(z0), imag(z0), "z0", 0.5, 0.5)
42     // Define the functions for the Steiner in- and circum-ellipse
43     deff("res=steinerInEllipseFunc(x, y)","res=Ai*x^2+Bi*x*y+Ci*y^2+Di*x+Ei*y+Fi")
44     deff("res=steinerCircumEllipseFunc(x, y)","res=Ac*x^2+Bc*x*y+Cc*y^2+Dc*x+Ec*y+Fc")
45     data = linspace(xmin - 5.0, xmax + 5.0, 1000)
46     x = data
47     y = data
48     // Display Steiner in-ellipse
49     contour2d(x, y, steinerInEllipseFunc, [0, 0], style=color("red"), axesflag=1)
50     // Display Steiner circum-ellipse
51     contour2d(x, y, steinerCircumEllipseFunc, [0, 0], style=color("blue"), axesflag=1)
52     // Display foci of the Steiner ellipses etc.
53     plot([real(zF1i), real(zF2i)], [imag(zF1i), imag(zF2i)], "ro")
54     plot([real(zF1c), real(zF2c)], [imag(zF1c), imag(zF2c)], "bo")
55     xstringb(real(zF1i), imag(zF1i), "zF1i", 0.5, 0.5)
56     xstringb(real(zF2i), imag(zF2i), "zF2i", 0.5, 0.5)
57     xstringb(real(zF1c), imag(zF1c), "zF1c", 0.5, 0.5)
58     xstringb(real(zF2c), imag(zF2c), "zF2c", 0.5, 0.5)
59     plotLine(zF1c, zF2c)
60  endfunction
```

**Listing 16 Function to plot the Steiner in- and circum-ellipse of the triangle $z_1$, $z_2$, $z_3$**

To really plot a triangle and its Steiner ellipses one has to execute the Scilab script ***steinerEllipses.sce*** (listing 17). The source code is nearly identical to that of ***mardensTheorem.sce*** (listing 2) except line 15. The result produced by running this script for the inputs **$z1 = 1 + \%i * 12$, $z2 = 4 - \%i * 2$** and **$z3 = -6$** can be seen in figure 7.

steinerEllipses.sce ✖

```
1  // Copyright (C) Klaus Rohe, D-85625 Glonn
2  clc()
3  format('v', 5);
4  lib("F:\CASTest\Scilab\MardensTheorem\mardenLib");
5  z1 = complex(1, 9.5);
6  z2 = complex(4, -1.5);
7  z3 = complex(-6, -1);
8  loopexit = %F;
9  while ~loopexit do
10     [z1, z2, z3, collinear, loopexit] = inputTrianglePoints(z1, z2, z3);
11     if loopexit then
12         printf("\nBye-bye!!\n");
13     else
14         if ~collinear then
15             plotSteinerEllipses(z1, z2, z3);
16         else
17             z1s = string(z1);
18             z2s = string(z2);
19             z3s = string(z3);
20             msg = msprintf("z1=%s, z2=%s z3=%s are collinear!", z1s, z2s, z3s);
21             messagebox(msg, "Collinearity", "info", "modal");
22         end
23     end
24 end
```

**Listing 17 Scilab script *steinerEllipses.sce* which has to be executed to plot the Steiner in- and circum-ellipse of a triangle $z_1$, $z_2$, $z_3$ to a graphics window**

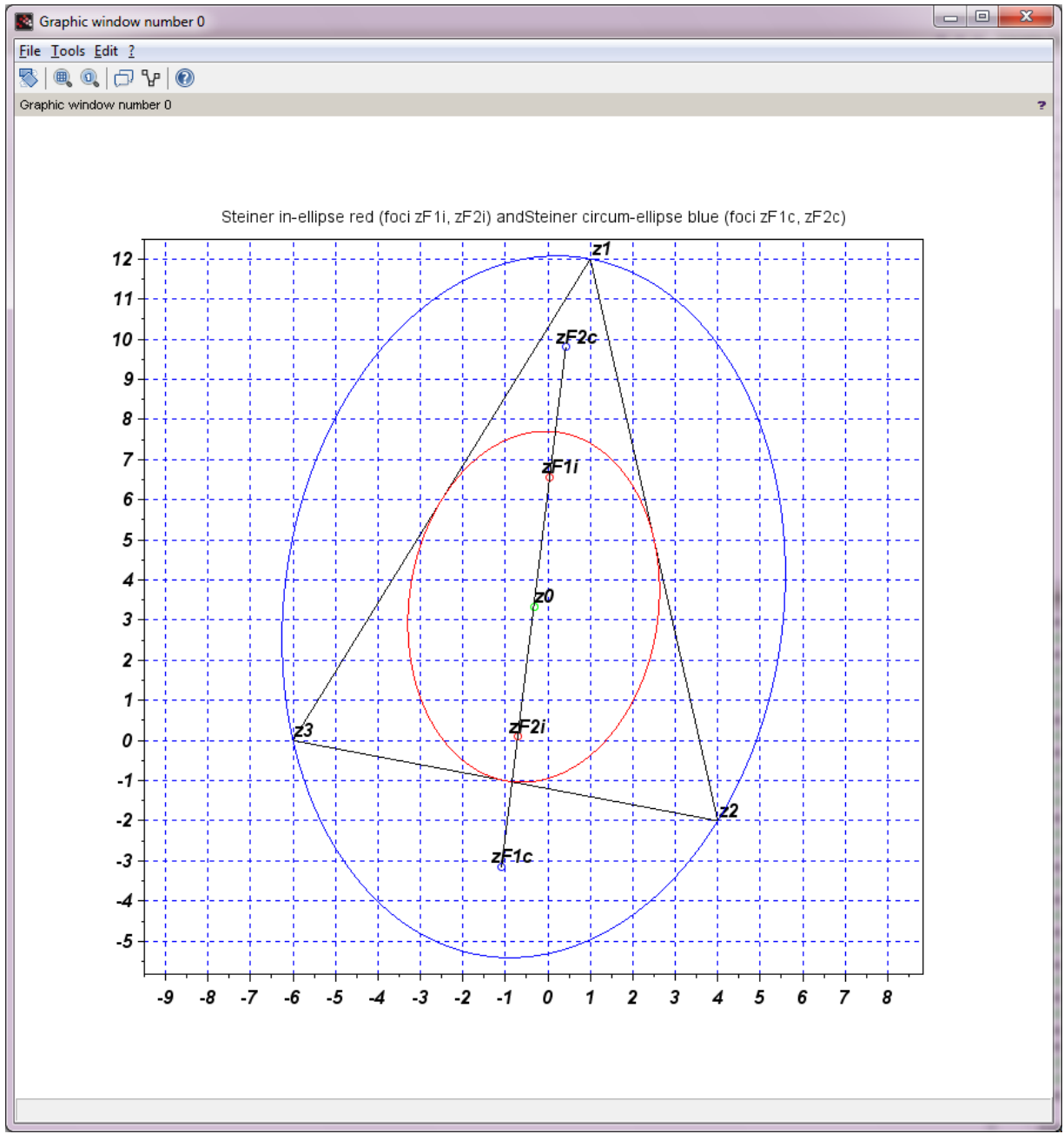

**Figure 7 Result of running the Scilab script *steinerEllipses.sce* for the inputs z1=1+%i*12, z2=4-%i*2 and z3=-6**

## (III) Organization of the source code

The source code of this Scilab application is organized as follows: there is a directory called
**MardensTheorem**. It contains the "main programs" **mardensTheorem.sce** and **steinerEllipses.sce**.
The folder **MardensTheorem** has a subdirectory with the name **mardenLib**. This directory contains all
the **\*.sci** files. The content of each of the file is the source code of one function. Table 1 contains an
alphabetically ordered list of these files. I have used the naming convention that the file has the
name of the function which it implements followed by the extension **.sci**. The files are compiled to a
Scilab library using the **genlib** command (see Scilab documentation).

**Table 1 List of Scilab functions presented and discussed above**

| | File name |
|---|---|
| 1 | buildTitleString.sci |
| 2 | computeAxis.sci |
| 3 | computeSteinerEllipses.sci |
| 4 | computeSteinerInEllipse.sci |
| 5 | computexySquare.sci |
| 6 | conicByFivePoints.sci |
| 7 | displayTriangleStuff.sci |
| 8 | initGraphics.sci |
| 9 | inputTrianglePoints.sci |
| 10 | plotLine.sci |
| 11 | plotSteinerEllipses.sci |
| 12 | plotSteinerInEllipse.sci |
| 13 | plotTriangle.sci |
| 14 | printEllipseParameters.sci |
| 15 | steinerInEllipsePoints.sci |

The sources of the Scilab functions and scripts can be obtained –with no warranty– by sending an
email to the author. The author is also grateful for hints concerning typos, errors and how the paper
can be improved.